# Observation of Moiré Excitons in $WSe_2/WS_2$ Heterostructure Superlattices

Chenhao Jin[1]†, Emma C. Regan[1,2]†, Aiming Yan[1,3], M. Iqbal Bakti Utama[1,4], Danqing Wang[1,2], Ying Qin[5], Sijie Yang[5], Zhiren Zheng[1], Kenji Watanabe[6], Takashi Taniguchi[6], Sefaattin Tongay[5], Alex Zettl[1,3,7], Feng Wang[1,3,7]*

[1] Department of Physics, University of California at Berkeley, Berkeley, California 94720, United States.

[2] Graduate Group in Applied Science and Technology, University of California at Berkeley, Berkeley, California 94720, United States

[3] Material Science Division, Lawrence Berkeley National Laboratory, Berkeley, California 94720, United States.

[4] Department of Materials Science and Engineering, University of California at Berkeley, Berkeley, California 94720, United States.

[5] School for Engineering of Matter, Transport and Energy, Arizona State University, Tempe, Arizona 85287, United States

[6] National Institute for Materials Science, 1-1 Namiki, Tsukuba, 305-0044, Japan.

[7] Kavli Energy NanoSciences Institute at University of California Berkeley and Lawrence Berkeley National Laboratory, Berkeley, California 94720, United States.

† These authors contributed equally to this work

* Correspondence to: fengwang76@berkeley.edu

**Abstract**:

Moiré superlattices provide a powerful tool to engineer novel quantum phenomena in two-dimensional (2D) heterostructures, where the interactions between the atomically thin layers qualitatively change the electronic band structure of the superlattice. For example, mini-Dirac points, tunable Mott insulator states, and the Hofstadter butterfly can emerge in different types of graphene/boron nitride moiré superlattices, while correlated insulating states and superconductivity have been reported in twisted bilayer graphene moiré superlattices[1-12]. In addition to their dramatic effects on the single particle states, moiré superlattices were recently predicted to host novel excited states, such as moiré exciton bands[13-15]. Here we report the first observation of moiré superlattice exciton states in nearly aligned $WSe_2/WS_2$ heterostructures. These moiré exciton states manifest as multiple emergent peaks around the original $WSe_2$ A exciton resonance in the absorption spectra, and they exhibit gate dependences that are distinctly different from that of the A exciton in $WSe_2$ monolayers and in large-twist-angle $WSe_2/WS_2$ heterostructures. The observed phenomena can be described by a theoretical model where the periodic moiré potential is much stronger than the exciton kinetic energy and creates multiple flat exciton minibands. The moiré exciton bands provide an attractive platform to explore and control novel excited state of matter, such as topological excitons and a correlated exciton Hubbard model, in transition metal dichalcogenides.

A moiré superlattice can form between two atomically thin materials with similar lattices, and its period varies continuously with the twist angle between the constituent layers. The periodic moiré pattern introduces a new length and energy scale, providing a powerful new way to control quantum phenomena in 2D heterostructures[1-12]. The most striking moiré superlattice phenomena emerge in the "strong-coupling" regime, where the periodic moiré potential dominates over the relevant kinetic energy and qualitatively changes the electronic band structure and the electron wavefunction in the heterostructure. Recently, it was reported that "strong-coupling" moiré superlattices can generate flat electronic bands, leading to exotic phases such as correlated insulating states and superconductivity in magic-twist-angle bilayer graphene and tunable Mott insulator states in trilayer graphene/boron nitride heterostructures[1-6].

Moiré superlattices also offer exciting opportunities to engineer the band structure of collective excitations, such as excitons in 2D semiconducting heterostructures. Monolayer transition metal dichalcogenides (TMDCs) are direct bandgap semiconductors that feature strong light-exciton interactions and dramatically enhanced electron-electron interactions. Exciton binding energies in monolayer TMDCs can be hundreds of meV – orders of magnitude larger than what is seen in typical semiconductors like silicon or GaAs[16,17] – which leads to well defined dispersive exciton bands in the Brillouin zone. Recently it was predicted that moiré superlattices in the "strong coupling" regime could lead to moiré exciton minibands in TMDC heterostructures[13-15], which are distinctly different from the separate electron and hole minibands due to the strong electron-hole correlation.

Here we report the first experimental observation of moiré excitons in nearly aligned $WSe_2/WS_2$ heterostructures. The moiré superlattice splits the $WSe_2$ A exciton resonance into multiple peaks that all exhibit comparable oscillator strengths in the absorption spectrum. Furthermore, the emergent exciton peaks show distinct doping dependences that are different from that of the A exciton in WSe$_2$ monolayers and in large-twisting-angle WSe$_2$/WS$_2$ heterostructures. This unusual behavior can be understood using an empirical model for moiré excitons with a peak-to-peak exciton moiré potential of 250 meV. The periodic potential energy is much larger than the exciton kinetic energy of 8 meV within the first mini-Brillouin zone, and it completely changes the exciton dispersion in the moiré superlattice, leading to flat low-energy exciton bands with highly localized exciton density of states. The near-aligned WSe$_2$/WS$_2$ moiré superlattice can therefore potentially host a variety of novel excitonic states, such as topological exciton bands and a strongly-correlated exciton Hubbard model[13-15,18,19].

Figure 1a and b show an optical microscopy image and a side-view schematic of a representative WSe$_2$/WS$_2$ heterostructure device (D1). The results measured from device D1 are reproducible in all near-aligned heterostructures that we fabricated (see supplementary information). The WSe$_2$/WS$_2$ heterostructure is encapsulated in thin hexagonal boron nitride (hBN) layers. Few layer graphite (FLG) flakes are used for both the bottom gate and the electrical contacts to the heterostructure. The carrier concentration in the heterostructure can be tuned continuously with the back gate voltage $V_g$. All of the two-dimensional materials were first mechanically exfoliated from bulk crystals and then stacked together by a dry transfer method using a polyethylene terephthalate (PET) stamp (see Methods). The whole stack was then transferred onto a 90 nm SiO$_2$/Si substrate. The relative twist angle between the WSe$_2$ and WS$_2$ layers was determined optically using polarization-dependent second harmonic generation (SHG) measurements. Characteristic six-fold SHG patterns are clearly observed for the WSe$_2$ layer (green color) and the WS$_2$ layer (yellow color) in Fig. 1c, from which we can determine the relative twist angle between the two layers to be <0.5 degree (see supplementary information).

For a near-zero twist angle heterostructure, the lattice mismatch between the two layers is dominated by the intrinsic lattice constant difference of about 4% (Ref. 20) , which leads to a moiré periodicity $L_M \approx 8$ nm (Fig. 1d). Similar moiré superlattices have been observed experimentally using scanning tunneling microscope (STM) in aligned WSe$_2$/MoS$_2$ heterostructures[21,22], which have lattice constants almost identical to the WSe$_2$/WS$_2$ heterostructure. To confirm the formation of the moiré superlattice in our WSe$_2$/WS$_2$ heterostructure, we prepared another device (D2) on a transmission electron microscopy (TEM) grid and collected high resolution TEM images of the device (see methods). Rich sets of diffraction patterns are observed in the Fourier transform of the TEM image in Fig. 1e, which shows the main diffraction peaks from the WSe$_2$, WS$_2$, and hBN layers as well as side peaks from the local reconstruction of the atomic structures due to layer-layer interactions. A zoomed-in image (Fig. 1f) shows a well-defined hexagonal lattice in the center region that corresponds to a real space periodicity of ~ 8 nm. This indicates that a periodic lattice distortion with ~ 8 nm periodicity exists in the heterostructure in real space, which is consistent with strong layer-layer interaction and significant lattice reconstruction within the moiré superlattice observed in previous STM studies[21,22].

We probe the moiré excitons in WSe$_2$ with optical spectroscopy at a temperature of 10 Kelvin. Figure 2a shows the photoluminescence (PL) spectrum of device D1 (blue curve) and a reference monolayer WSe$_2$ sample (green curve) in both linear (main panel) and logarithmic (inset) scale. The heterostructure PL features a single peak at 1.409eV, corresponding to the emission from the interlayer exciton, and does not show any emission from WSe$_2$ A exciton. This indicates an efficient interlayer charge transfer across the whole measured region that leads to strong quenching of the WSe$_2$ PL[23,24]. The exciton absorption in the same heterostructure region is directly probed through reflection contrast measurements (top panel in Fig. 2b), where a slowly varying background has been subtracted to better resolve the resonances (see supplementary information). The absorption spectrum from D1, a nearly aligned heterostructure, is strikingly different from that of a large-twist-angle $WSe_2/WS_2$ heterostructure measured at the same condition (lower panel in Fig. 2b). We focus on the spectral range between 1.6 to 1.8 eV as it is well-separated from all $WS_2$ resonances. While the large-twist-angle heterostructure shows only a $WSe_2$ A exciton peak at 1.715 eV, three prominent peaks emerge in device D1 at 1.683, 1.739, and 1.776 eV, labeled as resonance I, II, and III, respectively. All three resonances show strong absorption, with peak II and peak III having oscillator strengths at 20% and 50% of the peak I value. To better understand these new exciton peaks, we measured the photoluminescence excitation (PLE) spectrum of the device D1 (black dots in Fig. 2c) by monitoring the interlayer exciton emission intensity as the excitation photon energy was swept from 1.6 to 2.1 eV. The excitation spectrum shows perfect correspondence to the results from reflection spectroscopy (blue line in Fig. 2c). In particular, the new exciton peaks between 1.6 to 1.8 eV all give rise to strong enhancement of the interlayer exciton emission at 1.409 eV, indicating that they arise from the strongly coupled $WSe_2/WS_2$ heterostructure.

To further investigate the nature of the emergent exciton resonances, we measure their doping dependence (Fig. 3a). The horizontal and vertical axes represent the photon energy and gate voltage $V_g$, respectively, and the color corresponds to reflection contrast. The charge neutral point is approximately at $V_g=0$, and positive and negative $V_g$ values correspond to electron- and hole-doping, respectively. The three main peaks in the $WSe_2$ A exciton range show rich behavior, with dramatic spectral changes upon both electron and hole doping. The strong gate-dependence upon electron doping is particularly remarkable: Due to the type-II band alignment in $WSe_2/WS_2$ heterostructures, doped electrons reside mostly in the $WS_2$ layers and tend to have relatively weak effects on the intra-layer A exciton resonance in $WSe_2$ (Ref. 25,26). Indeed, previous studies of large-twist-angle $WSe_2/WS_2$ heterostructures shows that the $WSe_2$ A exciton resonance only experiences a slight redshift upon electron doping of the heterostructure[26]. In contrast, the exciton peaks in D1, a nearly aligned heterostructure with a large moiré superlattice, show unusual dependences on electron doping that varies for different peaks (Fig. 3b). Both peak I and peak III are strongly modified at increasing electron concentration: Peak I shows a strong blueshift and transfers its oscillator strength to another emergent peak at lower energy (I'), and peak III also shows a strong blueshift with diminished oscillator strength. On the other hand, peak II remains largely unchanged except for a small energy shift.

The strong effect of electrons in $WS_2$ on certain exciton transitions in $WSe_2$ indicates dramatically enhanced interlayer electron-exciton interactions through the moiré superlattice. In addition, the strikingly different gating behavior of the exciton peaks cannot be explained by any established electron-exciton interactions in monolayers, such as dielectric screening effects or trion formation, which affect all exciton peaks in a similar fashion[27-29]. Instead, it indicates that the exciton peaks I, II, and III correspond to very different exciton states within the moiré superlattice.

Both the emergence of multiple exciton peaks around the $WSe_2$ A exciton resonance and their peculiar electron doping dependence can be understood within an empirical theory in which a periodic moiré exciton potential in the "strong coupling" regime is introduced. We follow the theoretical model in Ref. 13 and describe the center-of-mass motion of $WSe_2$ A excitons using the Hamiltonian

$$H = H_0 + \sum_{j=1}^{6} V_j \exp(i\boldsymbol{b}_j \cdot \boldsymbol{r}), \quad (1)$$

where $H_0$ is the low energy effective Hamiltonian for the A exciton 1s state in monolayer WSe$_2$. $V_j$ describes the effective potential on the exciton created by the moiré pattern; its momentum is given by the reciprocal lattice vectors of the moiré superlattice, $\boldsymbol{b}_j$ (see supplementary information). Owing to the three-fold rotational symmetry and Hermitian requirement, only one component in $V_j$ is independent and can be defined as $V_1 = V\exp(i\psi)$.

The exciton dispersion in the mini-Brillouine Zone (mBZ) can be directly calculated from this model (see Fig. 4, a-c). Without the moiré potential, the exciton shows two continuous bands at low energy (Fig. 4a). These two bands are degenerate at $\gamma$ point, and have parabolic and linear dispersion, respectively, as a consequence of the intervalley exchange interaction[13,30]. Because photons have negligible momentum, only the lowest energy exciton can interact with light, giving a single strong peak at $E = E_0$ in the absorption spectrum (Fig. 4d). The moiré potential can mix exciton states with momenta that differ by $\boldsymbol{b}_j$, leading to additional absorption peaks from the $\gamma$ point states of higher-energy minibands.

When the moiré potential is weak ($V = 5$ meV, Fig. 4b), the exciton dispersion remains largely unchanged. Therefore, the emergent side peak in absorption always appears at ~30 meV above the main peak, regardless of the exact form of the moiré potential (Fig. 4e). Furthermore, the amplitude of the side peak is orders of magnitude smaller than the main peak due to the weak mixing between states. These features pose sharp contrast to the experimental absorption spectrum and cannot explain our observations. On the other hand, a larger moiré potential that corresponds to the "strong coupling" regime dramatically modifies the exciton dispersion, (see

Fig. 4c). As a result, the energy of the moiré exciton states in different minibands (labeled I to III in Fig. 4c), as well as of the corresponding absorption peaks (peak I to III in Fig. 4f), become sensitively dependent on the moiré potential. In addition, the strong mixing between different exciton states make their oscillator strengths comparable to each other. By taking $V = 25$ meV and $\psi = 15°$, the simulated absorption spectrum can reproduce our experimental observation (Fig. 4f, see also supplementary information). This moiré potential has a peak-to-peak amplitude of ~ 250 meV, which is much larger than the exciton kinetic energy of ~ 8 meV within the first mini-Brillouin zone (see supplementary information).

The dramatic change in the exciton dispersion in momentum space implies that the exciton center-of-mass wavefunction is also strongly modified in real space. Figures 4g-i show the distribution of the exciton probability density for states I to III in the moiré superlattice. The originally homogeneous wavefunction distribution is dramatically changed by the moiré potential. For example, the lowest energy state (state I) is trapped around the moiré potential minimum (labeled as point α) in a length scale much smaller than the moiré superlattice (Fig. 4g). As a result, excitons in different moiré periods are well separated, forming an effective exciton lattice with significantly reduced hopping between neighboring lattice sites, which is consistent with the significantly reduced bandwidth in their momentum dispersion.

The peculiar wavefunction distribution of moiré excitons in the "strong coupling" regime introduces a new degree of freedom that is determined by the exciton location in the moiré superlattice. Interestingly, both peak I and III are centering around the same point α, while peak II has its largest amplitude at a different point β (Fig. 4, h-i). The difference in real space position between moiré exciton states can account for their distinctive doping-dependence: The doped electrons will also have localized density of states in real space[21,22]. If the gate-induced electrons in $WS_2$ are also localized at point α in the moiré superlattice, they will predominantly change the exciton peaks I and III and leave exciton peak II little affected, as observed in the experiment.

We note that a complete description of the moiré exciton optical spectra will require a much more sophisticated model that fully accounts for the lattice relaxation and corrugation, as well as the interlayer electronic states hybridization in the heterostructure moiré superlattice, which is beyond the scope of this study. Nevertheless, our simple moiré exciton model captures most of the salient features observed experimentally, and it shows that $WS_2/WSe_2$ heterostructures exhibit sufficiently strong interlayer interaction to enter the "strong coupling" regime for excitons, where the moiré excitons become spatially concentrated at well-separated points and form a quantum array in an extended moiré superlattice[13-15]. The significantly reduced exciton bandwidth also makes this artificial exciton lattice a promising platform for realizing exotic phases such as a topological exciton insulator and a strongly-correlated exciton Hubbard model.

**Methods**:

Heterostructure preparation for optical measurements: WSe$_2$/WS$_2$ heterostructures were fabricated using a dry transfer method with a polyethylene terephthalate (PET) stamp[31]. Monolayer WSe$_2$, monolayer WS$_2$, few-layer graphene, and thin hBN flakes were exfoliated onto silicon substrates with a 90 nm SiO$_2$ layer. Polarization-dependent SHG was used to determine the relative angle between the WS$_2$ and WSe$_2$ flakes (see text and supplementary information for details). A PET stamp was used to pick up the top hBN flake, the WS$_2$ monolayer, the WSe$_2$ monolayer, several few-layer graphene flakes for electrodes, the bottom hBN flake, and the few-layer graphene back gate in sequence. The angle of the PET stamp was adjusted between picking up the WS$_2$ and the WSe$_2$ to assure a near-zero twist angle between the flakes. The PET stamp with the above heterostructure was then stamped onto a clean Si substrate with 90 nm SiO$_2$, and the PET was dissolved in dichloromethane for 12 hours at room temperature. The PET and samples were heated to 60 °C during the pick-up steps and to 130 °C for the final stamp process. Contacts (~75 nm gold with ~5 nm chromium adhesion layer) to the few-layer graphene flakes were made using electron-beam lithography and electron-beam evaporation.

Heterostructure preparation for TEM: WS$_2$/WSe$_2$ heterostructures were prepared for TEM characterization using a modified dry transfer technique with a PET stamp. WS$_2$ monolayers, WSe$_2$ monolayers, and thin hBN flakes were exfoliated and SHG measurements were used to determine flake orientation, as described above. A PET stamp was used to pick up the top hBN flake, the WS$_2$ monolayer, the WSe$_2$ monolayer, and the bottom hBN flake in sequence. A Ted Pella Quantifoil TEM grid with 2 µm holes (657-200-AU) was placed on a silicon chip that was attached to the transfer stage. The PET stamp was lowered until it was in contact with the TEM grid, and then the temperature was raised to 80 °C until the stamp and the grid were well contacted, as seen through an optical microscope. The PET stamp and TEM grid were then placed in dichloromethane for 12 hours at room temperature to dissolve the PET.

High-resolution TEM imaging and FFT analysis: High-resolution TEM images of the hBN-encapsulated $WS_2/WSe_2$ heterostructure were taken under 200 keV accelerating voltage for the electron beam. A fast Fourier transform with high-pass filter was performed on each high-resolution TEM image to show the superlattice periodicity in the $WS_2/WSe_2$ heterostructure in reciprocal space.

**Data availability.**

The data that support the findings of this study are available from the corresponding author upon reasonable request.

**Acknowledgements**:

This work was supported primarily by the Director, Office of Science, Office of Basic Energy Sciences, Materials Sciences and Engineering Division of the U.S. Department of Energy under contract no. DE-AC02-05-CH11231 (van der Waals heterostructures program, KCWF16). The device fabrication was supported by the NSF EFRI program (EFMA-1542741). PLE spectroscopy of the heterostructure is supported by the US Army Research Office under MURI award W911NF-17-1-0312. Growth of hexagonal boron nitride crystals was supported by the Elemental Strategy Initiative conducted by the MEXT, Japan and JSPS KAKENHI Grant

Numbers JP15K21722. S.T. acknowledges the support from NSF DMR 1552220 NSF CAREER award for the growth of WS2 and WSe2 crystals.

**Author contributions:** C.J. and E.C.R. contributed equally to this work. F.W. and C.J. conceived the research. C.J. and E.C.R. carried out optical measurements. A.Y. and A.Z. performed TEM measurements. C.J., F.W., E.C.R. and D.W. performed theoretical analysis. E.C.R., M.I.B.U., D.W. and Z.Z. fabricated van der Waals heterostructures. Y.Q., S.Y. and S.T. grew WSe2 and WS2 crystals. K.W. and T.T. grew hBN crystals. All authors discussed the results and wrote the manuscript.

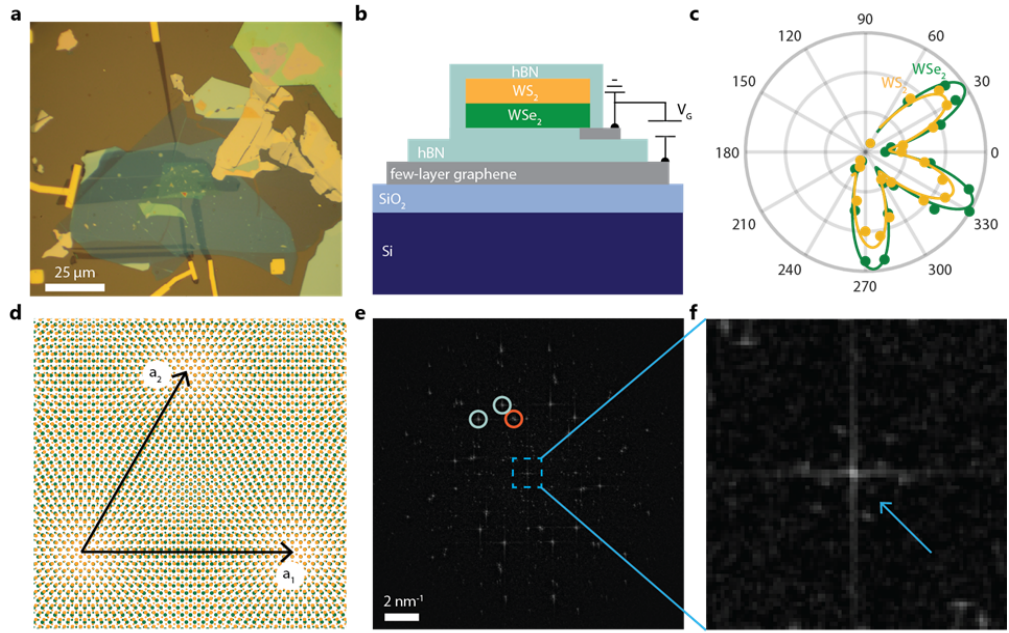

**Figure 1 | Moiré superlattice in near-zero twist angle WSe$_2$/WS$_2$ heterostructure. a, b,** Optical microscope image (**a**) and side-view illustration (**b**) of a representative near-zero twist angle heterostructure (device D1). **c,** The polarization-dependent SHG signal measured on the monolayer WSe$_2$ (green circles) and WS$_2$ (yellow circles) regions in device D1 and the corresponding fittings (green and yellow curves). The SHG results confirm that the WSe$_2$ and WS$_2$ twist angle is smaller than our experimental uncertainty of 0.5 degree. **d,** Illustration of the moiré superlattice in real space. The superlattice vectors, a$_1$ and a$_2$, have a length of ~ 8 nm. **e, f,** Fourier transform of the TEM image of another near-zero twist angle WSe$_2$/WS$_2$ heterostructure (device D2) (**e**) and the zoom-in plot at the center region (**f**). Representative first order diffraction points are labelled by circles in **e** for top and bottom hBN (light blue) and the WSe$_2$/WS$_2$ heterostructure (red), respectively. Two well-defined hexagonal lattices are observed in the center region, and the inner one (arrow in **f**) corresponds to a periodic lattice distortion with ~ 8 nm periodicity, consistent with the formation of a moiré superlattice. Scale bar: 2 nm$^{-1}$.

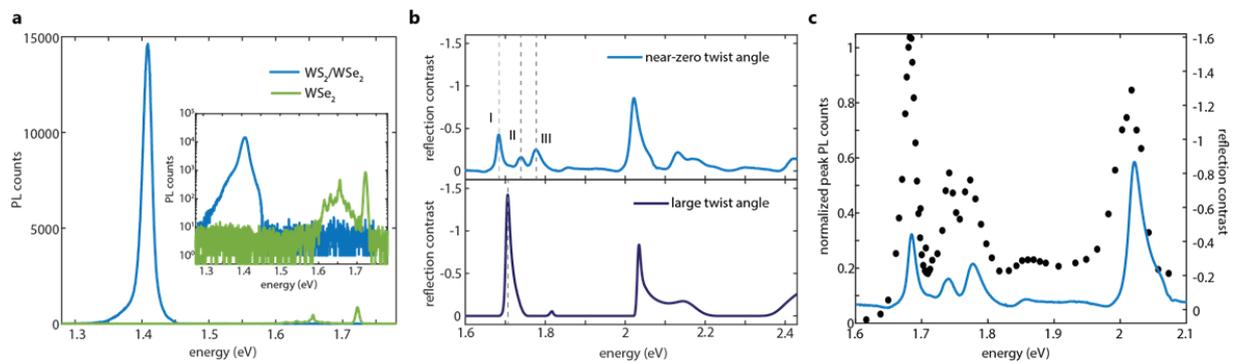

**Figure 2 │ Moiré exciton states in WSe$_2$/WS$_2$ moiré superlattice. a,** PL spectrum of device D1 (blue) and a reference monolayer WSe$_2$ sample (green curve) in both linear (main panel) and logarithmic (inset) scale. The complete disappearance of monolayer PL in the heterostructure indicates efficient interlayer coupling across the whole measured region. **b,** Reflection contrast spectrum of device D1 (blue color, upper panel) compared to a large-twist-angle WSe$_2$/WS$_2$ heterostructure device (navy color, lower panel). The latter only shows a single resonance in the energy range between 1.6 to 1.8 eV from the WSe$_2$ A exciton state. In contrast, the moiré superlattice formed in device D1 gives rise to three prominent peaks with comparable oscillator strength in this range (labelled as I to III), corresponding to different moiré exciton states. **c,** Comparison between the interlayer exciton photoluminescence excitation spectrum (black dots) and the reflection spectrum (blue curve). Strong enhancement of interlayer exciton photoluminescence is observed when excited at all moiré exciton states, indicating that all states are from the strongly-coupled WSe$_2$/WS$_2$ heterostructure.

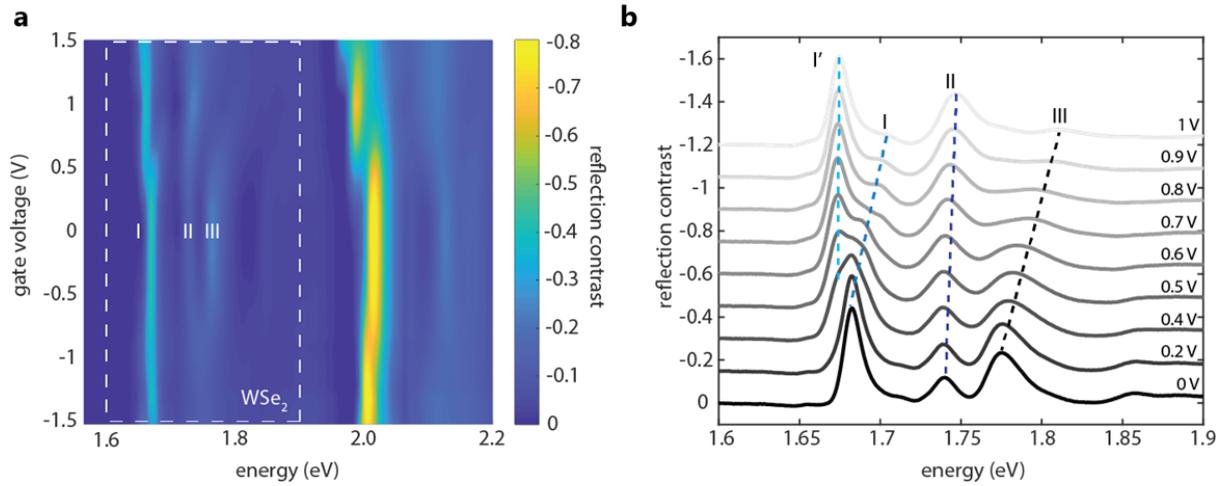

**Figure 3 | Doping dependence of the moiré exciton resonances. a,** Gate-dependent reflection contrast spectrum of device D1 with both electron- (positive $V_g$) and hole- (negative $V_g$) doping. White dashed box encloses the photon energy range near the $WSe_2$ A exciton, where the three prominent moiré exciton states (labeled as I, II, and III) appear. **b,** Detailed reflection contrast spectra in the $WSe_2$ A exciton range on the electron-doping side, which reveal unusual gate-dependence of the moiré exciton states: Peak I shows a strong blueshift and transfers its oscillator strength to another emerging peak at lower energy (I'), and peak III shows a strong blueshift with diminished oscillator strength. On the other hand, peak II remains largely unchanged except for a small energy shift. These observations cannot be explained by any established electron-exciton interactions in TMD monolayers.

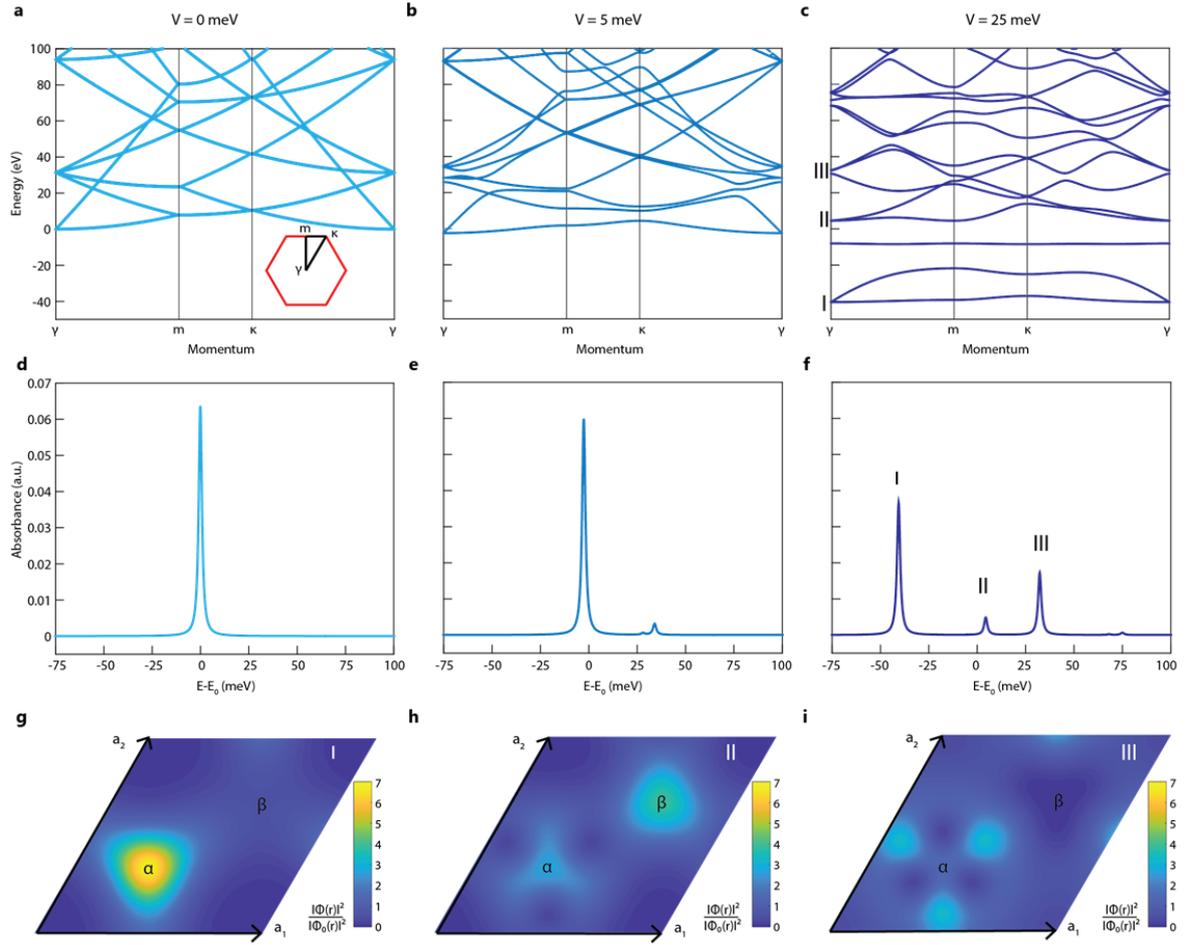

**Figure 4 | Moiré excitons in the "strong-coupling" regime. a-c,** WSe$_2$ A exciton dispersion in the mini-Brillouin zone with a moiré potential parameter V=0 meV (zero coupling, **a**), 5meV (weak couping, **b**), and 25 meV (strong coupling, **c**), and the corresponding absorption spectrum (**d-f**). A broadening of 2 meV is used in calculating the absorption spectrum. Inset in (**a**) illustrates the mini-Brillouin zone in momentum space and the high-symmetry points. The absorption spectrum features a single resonance at energy $E_0$ at zero moiré potential (**d**), and shows a small side peak fixed at ~ 30 meV under a weak moiré potential (**e**). These features cannot explain our experimental observation. On the other hand, the exciton dispersion is strongly modified in the "strong-coupling" regime due to the strong mixing between different exciton states (**c**), which gives rise to multiple moiré exciton peaks with comparable oscillator

strength in the absorption spectrum (peak I to III in **f**) from different moiré mini-bands (state I to III in **c**). The experimentally observed reflection contrast can be reproduced by taking $V = 25$ meV and $\psi = 15°$. **g-i,** Real space distribution of exciton center-of-mass wavefunction in the "strong-coupling" regime. The strong moiré potential traps the lowest-energy exciton state I around its minimum point α (**g**). Interestingly, state III is also centered at point α, but state II is centered at a different point (**h, i**), which can account for the remarkably different gate dependence between the moiré exciton states.

**Supplementary Materials for**

**Observation of Moiré Excitons in WSe$_2$/WS$_2$ Heterostructure Superlattices**

Chenhao Jin†, Emma C. Regan†, Aiming Yan, M. Iqbal Bakti Utama, Danqing Wang, Ying Qin, Sijie Yang, Zhiren Zheng, Kenji Watanabe, Takashi Taniguchi, Sefaattin Tongay, Alex Zettl, Feng Wang*

† These authors contributed equally to this work

* Correspondence to: fengwang76@berkeley.edu

**1. Results from additional near-aligned heterostructures**

**2. Determination of the relative twist angle between WSe$_2$ and WS$_2$ layers**

**3. Subtraction of background in reflection contrast spectra**

**4. Dependence of moiré exciton absorption spectra on the moiré potential**

**5. Spatially localized exciton center-of-mass wavefunctions**

## 1. Results from additional near-aligned heterostructures

The moiré excitons observed in device D1 and described in the text are reproducible in all near-aligned heterostructures that we fabricated, with crystals from several commercial and academic sources. For example, Fig. S1 shows the reflection contrast and PLE measurement results from another near-aligned heterostructure device D3 when it is slightly n-doped. Four resonances (I', I, II, III) are clearly observed between 1.6 and 1.9 eV in both reflection contrast and PLE spectra, whose energy and amplitude match well with the corresponding resonances observed for device D1.

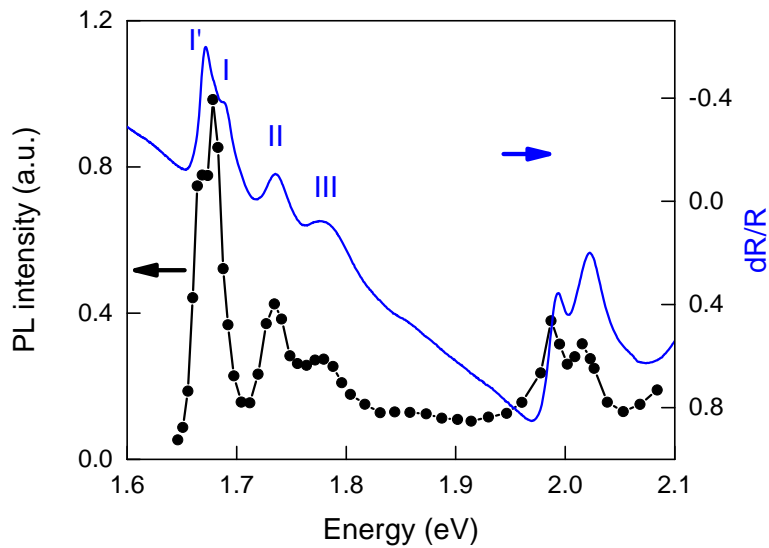

Fig. S1. Reflection contrast (blue) and PLE (black) spectra of another near-aligned heterostructure D3 at slight n-doping.

## 2. Determination of the relative twist angle between $WSe_2$ and $WS_2$ layers

The crystal orientation of $WSe_2$ and $WS_2$ flakes can be obtained from the second harmonic generation (SHG) polarization dependence. However, since the SHG patterns of both materials have six-fold rotational symmetry, the case of AA stacking (~ 0 twist angle) and AB stacking (~ 60 twist angle) cannot be differentiated by separately measuring the SHG of each material. On the other hand, the two cases can be distinguished by directly measuring the SHG of the heterostructure: in AA (AB) stacking case, the second harmonic field of the two layers will constructively (destructively) interfere, giving SHG signal stronger (weaker) than monolayers. Figure S2 shows the SHG intensity from $WSe_2$ alone, $WS_2$ alone, and heterostructure regions in device D1 measured with a 900 nm incident beam and the same experimental configuration. The $WSe_2$ and $WS_2$ regions show similar SHG intensity, while the heterostructure region show SHG intensity approximately four times larger than the monolayer. This result indicates that the twist angle between $WSe_2$ and $WS_2$ layers is near zero, i.e. AA-stacking.

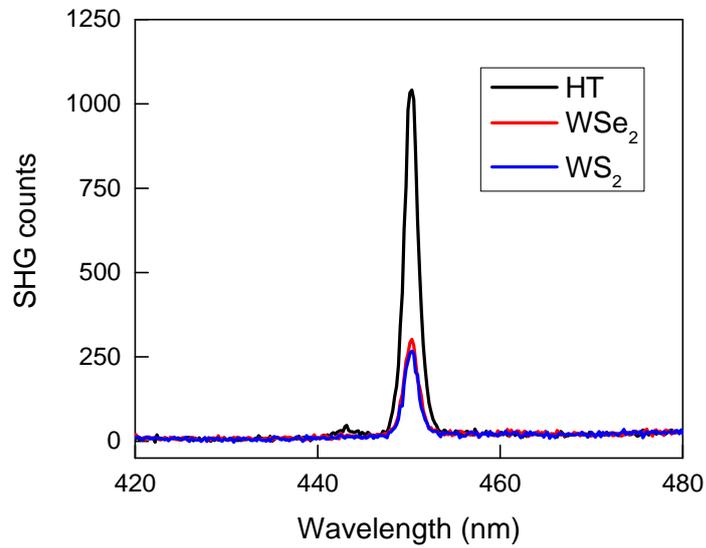

Fig. S2. SHG signal of device D1 measured on the $WSe_2$ alone (red), $WS_2$ alone (blue), and heterostructure (black) regions with the same experimental configuration.

## 3. Subtraction of background in reflection contrast spectra

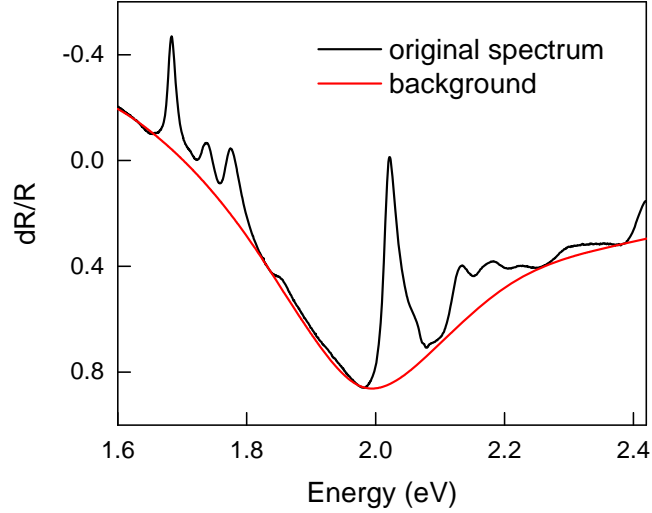

Fig. S3. Original reflection contrast (black) and the slow-varying background (red) obtained from polynomial fitting.

The $WSe_2/WS_2$ heterostructure is encapsulated in two pieces of hBN and placed on 90 nm $SiO_2$/Si substrate. This multi-film structure introduces interference between reflections at different interfaces, leading to a background signal from the real part of the dielectric function of the heterostructure[32], see Fig. S3. Because the phase difference between the multi-reflections in the interference depend on the light wavelength, this background signal will vary with photon energy. To better resolve the exciton absorption resonances (i.e. the imaginary part of dielectric function), we obtain a slowly-varying background signal through polynomial fitting of the spectra using regions away from resonances (red curve in Fig. S3). The same background is used universally to obtain background-subtracted spectra at charge neutral (Fig. 2 in text) and at different doping levels (Fig. 3 in text). The validity of the background subtraction is also confirmed by the comparison to PLE spectrum (Fig. 2c in text) because the latter only depends on the imaginary part of dielectric function and is background free.

## 4. Dependence of moiré exciton absorption spectra on the moiré potential

The low-energy effective Hamiltonian of the A-exciton in monolayer WSe$_2$ is described by

$$H_0 = \left(E_0 + \frac{\hbar^2 \mathbf{Q}^2}{2M}\right)\tau_0 + J|\mathbf{Q}|\tau_0 + J|\mathbf{Q}|[\cos(2\phi_\mathbf{Q})\tau_x + \sin(2\phi_\mathbf{Q})\tau_y],$$

where $\mathbf{Q}$ is exciton total momentum, $\phi_\mathbf{Q}$ is the polar angle of $\mathbf{Q}$ in the momentum space, $M \approx m_0$ is the total mass of the electron and the hole, $J = 0.04\text{eV} \cdot \text{nm}$ describes the intra- and inter-valley exchange interaction, and $\tau_j$ ($j = 0, x, y, z$) is the Pauli matrices for valley pseudospin[13]. Without the moiré potential, only the $\mathbf{Q} = 0$ exciton is bright due to the negligible momentum of photons, which has energy $E = E_0$.

With the moiré potential, the total Hamiltonian of the exciton becomes:

$$H = H_0 + V_m = H_0 + \sum_{j=1}^{6} V_j \exp(i\mathbf{b}_j \cdot \mathbf{r}).$$

When $V$ is small (i.e. the "weak coupling" regime), the effect of the moiré potential can be intuitively understood from perturbation theory. The first order correction to the wavefunction dictates that states at $\mathbf{Q} = \mathbf{b}_j$ will be mixed with the $\mathbf{Q} = 0$ state in the following way:

$$\Phi(\mathbf{b}_j)^1 - \Phi(\mathbf{b}_j)^0 = \frac{\langle \Phi(\mathbf{b}_j)^0 | V_m | \Phi(0)^0 \rangle}{E(\mathbf{b}_j) - E_0} \Phi(0)^0 \sim \frac{V_j}{4E_m} \Phi(0)^0,$$

where $\Phi(\mathbf{Q})$ is the exciton center-of-mass wavefunction at momentum $\mathbf{Q}$ and $E_m = \hbar^2 b_j^2/(8M)$ is the exciton kinetic energy within the first mini-Brillouin zone. $E_m \sim 8$ meV in a WSe$_2$/WS$_2$ moiré superlattice with periodicity of $\sim 8$ nm. Because the wavefunction of excitons at $\mathbf{Q} = \mathbf{b}_j$ now contains part of the bright exciton wavefunction, their transition from the ground state is no longer completely forbidden and has oscillator strength $\sim |V_j/(4E_m)|^2 = [V/(4E_m)]^2$, giving an additional absorption peak at $4E_m = 30$ meV above the main peak. The magnitude of the moiré potential can therefore be obtained by examining the amplitude of this side peak in the absorption spectrum.

Figure S4a shows the simulated exciton absorption spectra with phase $\psi = 15°$ and different magnitudes of the moiré potential. The side peak amplitude shows a monotonic increase with larger $V$. The quantitative dependence deviates from the square scaling law at large $V$, which is expected since the perturbation treatment fails in the "strong-coupling" regime. By comparing the experimental results to the simulation, we can extract the magnitude of the moiré potential to be $V \sim 25$ meV.

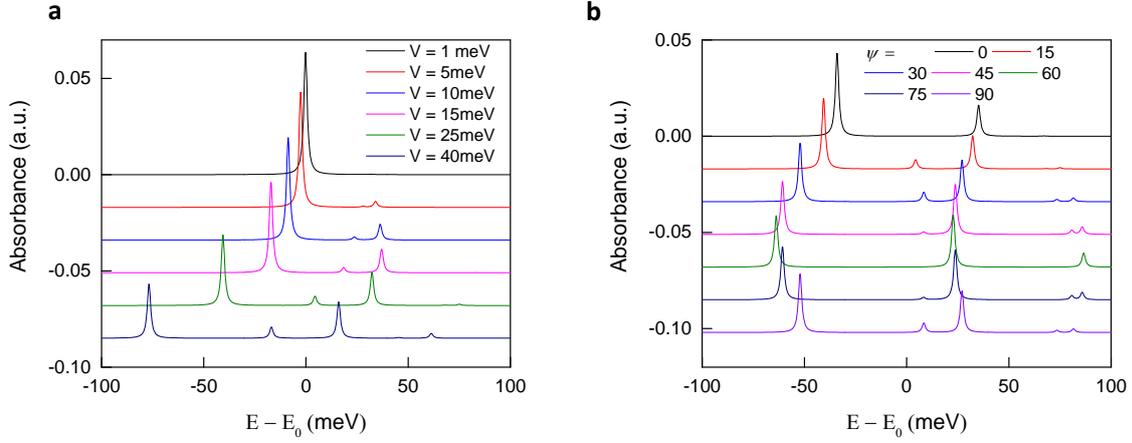

Fig. S4. **a**, Simulated exciton absorption spectra with different moiré potential magnitude $V$. $\psi = 15°$. **b**, Simulated exciton absorption spectra with different moiré potential phase $\psi$ and $V = 25$ meV. Different curves are vertically shifted for visual clarity.

The phase $\psi$ of the moiré potential plays a more subtle role: It has negligible effect on the absorption spectra in the "weak coupling" regime but will affect the spectra in the "strong coupling" regime, where higher order mixing between exciton states become important and different mixing paths start to interfere, as shown in Fig. S4b. The complicated dependence of the spectra on $\psi$ makes it difficult to have an accurate determination through comparison to the experiment. However, different $\psi$ will not qualitative change the properties of the system, e.g. between "weak coupling" regime and "strong coupling" regime.

### 5. Spatially localized exciton center-of-mass wavefunction

Figure S5 shows the real-space distribution of the moiré potential using the parameters $V = 25$ meV and $\psi = 15°$. After summing up the six components, the peak-to-peak amplitude of the moiré potential reaches ~ 250 meV, which is much larger than $E_m$. As a result, the lowest energy excitons are trapped around the potential minimum point (labelled as α), as discussed in the text.

We note that the exciton wavefunction discussed here refers to the center-of-mass envelop function for 1s exciton state, which is spatially homogeneous without the moiré superlattice. This should not be confused with the relative motion between the electron and the hole within an exciton that defines atom-like levels, e.g. 1s and 2p states.

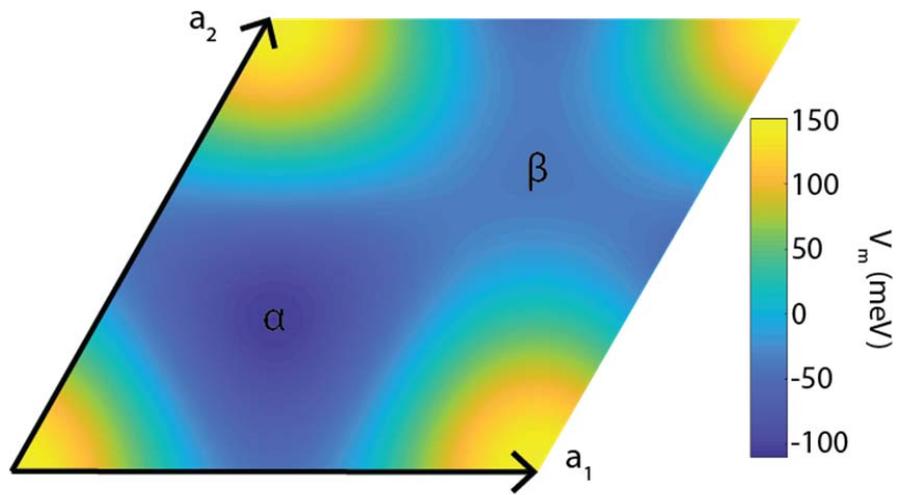

Fig. S5. Real space distribution of the moiré potential with $V = 25$ meV and $\psi = 15°$. The potential minimum is labelled as α.